\documentclass{article}
\usepackage{graphicx}

\PassOptionsToPackage{numbers, compress}{natbib}

\usepackage[final]{neurips_2024}

\usepackage[utf8]{inputenc} 
\usepackage[T1]{fontenc}    
\usepackage{hyperref}       
\usepackage{url}            
\usepackage{booktabs}       
\usepackage{amsfonts}       
\usepackage{nicefrac}       
\usepackage{microtype}      
\usepackage{xcolor}         

\title{AI TrackMate: Finally, Someone Who Will Give Your Music More Than Just ``Sounds Great!''}

\author{%
  Yi-Lin Jiang$^{1*}$ \quad Chia-Ho Hsiung$^{1*}$ \\
  \quad \textbf{Yen-Tung Yeh}$^2$ \quad \textbf{Lu-Rong Chen}$^1$ \quad \textbf{Bo-Yu Chen}$^1$ \\
  $^1$Rhythm Culture Corporation, Taiwan \\\quad $^2$Graduate Institute of Communication Engineering, National Taiwan University, Taiwan \\
  \texttt{\{rogerylc, chiahohsiung\}@gmail.com}\\
  \texttt{r12942179@ntu.edu.tw}, \texttt{\{warconvict, bernie40916\}@gmail.com}\\
}

\begin{document}

\maketitle

\begin{abstract}

The rise of \emph{``bedroom producers''} has democratized music creation, while challenging producers to objectively evaluate their work. To address this, we present AI TrackMate, an LLM-based music chatbot designed to provide constructive feedback on music productions. By combining LLMs' inherent musical knowledge with direct audio track analysis, AI TrackMate offers production-specific insights, distinguishing it from text-only approaches. Our framework integrates a Music Analysis Module, an LLM-Readable Music Report, and Music Production-Oriented Feedback Instruction, creating a plug-and-play, training-free system compatible with various LLMs and adaptable to future advancements. We demonstrate AI TrackMate's capabilities through an interactive web interface and present findings from a pilot study with a music producer. By bridging AI capabilities with the needs of independent producers, AI TrackMate offers on-demand analytical feedback, potentially supporting the creative process and skill development in music production. This system addresses the growing demand for objective self-assessment tools in the evolving landscape of independent music production.

\end{abstract}

\section{Introduction}

The advancement of music technology has revolutionized music production, consolidating various roles into the \emph{``bedroom producer''} \cite{Walzer2021, Darron2017, Paul2023}. This shift has democratized music creation, allowing individuals with limited musical background to produce finished pieces\cite{Walzer2021, Darron2017}. However, this solitary production environment presents challenges, particularly in objectively evaluating one's work \cite{Darron2017} and developing critical listening skills \cite{Walzer2021}. 

The lack of regular and objective feedback in isolated production environments threatens to limit the potential of many talented independent producers \cite{Darron2017}. Traditionally, these skills are developed in higher education or professional studio settings \cite{Walzer2021}, which may be inaccessible to many bedroom producers. Recent advances in AI technology, particularly Large Language Models (LLMs), offer a promising solution to this dilemma.

LLMs have demonstrated impressive capabilities in various domains \cite{Qingyun2023}, including music. Recent works have used LLMs for music captioning \cite{SeungHeon2023}, understanding, and reasoning \cite{Josh2023, Ruibin2024}. However, these approaches face limitations when applied to providing constructive feedback to independent music producers. They often lack specialized production knowledge \cite{Josh2023} or focus more on music theory with symbolic representation than production with audio track \cite{Ruibin2024}, with limited utility for producers working primarily with audio tracks.

To address these limitations, we propose AI TrackMate, an LLM-based music chatbot designed to provide objective and constructive feedback on music productions. Unlike previous approaches that require the training of new models \cite{SeungHeon2023} or the fine-tuning of existing ones \cite{Josh2023, Ruibin2024}, AI TrackMate leverages the inherent musical knowledge of LLMs \cite{Qixin2024}, focusing on providing detailed track information and guiding the LLM to think like a music producer.

Our framework consists of three key components: \emph{Music Analysis Module}, \emph{LLM Readable Music Report}, and \emph{Music Production-Oriented Feedback Instruction}. This design creates a plug-and-play, training-free system compatible with any LLM \cite{Qixin2024}, including those fine-tuned on open-source models \cite{Josh2023, Ruibin2024} and future iterations with enhanced music understanding capabilities. We demonstrate AI TrackMate through an interactive web-based interface, allowing music producers to easily upload their tracks and receive structured analytical feedback based on multiple musical aspects, supporting their production decision-making process. Moreover, we conducted a pilot study evaluating the system's effectiveness, involving an in-depth qualitative interview with a real music producer. 

AI TrackMate represents a significant step forward in AI-assisted music production, providing detailed analytical feedback on submitted music tracks. By bridging the gap between AI capabilities and the specific needs of independent music producers, AI TrackMate has the potential to significantly enhance the creative process and accelerate skill development in the evolving landscape of music production. Our interactive demo can be checked in \url{https://worzpro.github.io/aitrackmate-demo-page}, providing hands-on experience with comprehensive track analysis and insights into its real-world application and impact.

\section{System}
\label{system}

\begin{figure}[h]
    \centering
    \includegraphics[width=\linewidth]{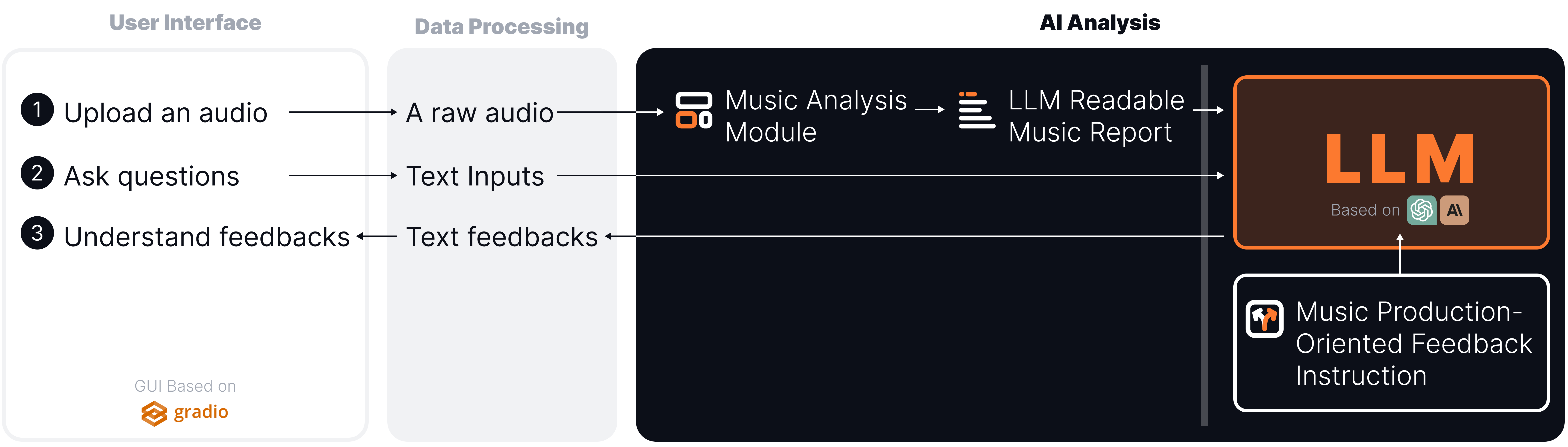}
    \caption{The system comprises three layers: (1) \ \textbf{User Interface} for audio upload, query input, and feedback reception; (2)  \textbf{Data Processing} for handling raw audio and text; and (3) \textbf{AI Analysis}, featuring a Music Analysis Module that transforms raw audio into LLM-readable report, and an LLM that processes these reports along with user queries. Guided by music production-oriented feedback instructions, the LLM generates insights comparable to those of a music producer.}
    \label{fig:system}
\end{figure}

Our system delivers AI-generated feedback through four core components, as illustrated in Figure \ref{fig:system}: \emph{Music Analysis Module}, \emph{LLM Readable Report}, \emph{Music Production-Oriented Feedback Instruction}, and \emph{Web-based UI}. The Music Analysis Module examines audio input across multiple aspects, mimicking a producer's detailed ear. LLM Readable Report structures this analysis, similar to how producers organize their thoughts before giving feedback. Music Production-Oriented Feedback Instructions guide the LLM, akin to a seasoned producer drawing on experience to form critiques. These components integrate into a user-friendly web UI for easy interaction. In the following section, we will explain the methodology and functionality of each component in detail, demonstrating how they work together to provide comprehensive music analysis and production quality feedback.

\subsection{Music Analysis Module}
\label{sec:music_analysis}
Rhythm, harmony and sound design constitute the core elements of music production. To facilitate an LLM's comprehension of these aspects, we conduct a comprehensive analysis of each element. For rhythm, this analysis encompasses onset detection, beat-and-downbeat tracking, and tempo estimation. With respect to harmony, we concentrate on key classification and chord recognition. In terms of sound design, we perform instrument recognition and extract timbral characteristics. In addition, to allow an LLM to grasp the expressive and subjective dimensions of music, we implement genre, theme, and emotion classification. Recognizing that most tracks exhibit diverse emotions and instruments across various sections, we extend our predictions of emotion and instrumentation beyond the track level to include structure-level analysis.

We employ All-In-One \cite{taejun2023allinone} for detecting beats and downbeats, estimating tempo, and segmenting structure. We utilize Madmom \cite{madmom} for onset detection and key classification, and autochord \cite{Bayron2021} for chord recognition. To extract timbral characteristics, we rely on timbral models by \textit{AudioCommons} \footnote{\url{https://github.com/AudioCommons/timbral_models}}. Additionally, Essentia \cite{Bogdanov2013} is leveraged for instrument recognition, theme classification, and emotion classification.

\subsection{LLM Readable Music Report}

We implement an iterative refinement process to optimize LLMs' interpretation of music analysis results, ensuring alignment with musicians' needs. This process commences by feeding raw meta-data (including genre, chords, rhythm, and emotion metrics) directly to the LLM. Initial interpretations are typically suboptimal due to the data's complexity. For example, in chord analysis, raw data might include time-stamped chord labels (e.g., ``end'': 4.37, ``label'': ``G:min'', ``start'': 0.79). The LLM's initial analysis often identifies dominant chords and general texture but lacks depth. Subsequently, we augment the data based on identified gaps by calculating additional statistical metrics. In our chord analysis example, we introduce metrics such as total chord changes, dominant chord identification, and major/minor chord counts. This refinement continues for 2-3 iterations, each enhancing the LLM's understanding. By the final iteration, we typically incorporate more nuanced metrics such as average chord duration and common progressions. This enables the LLM to generate comprehensive insights, including analyses of chord usage patterns, transition pacing, and emotional complexity. The process concludes with a secondary LLM that evaluates the output of each iteration for clarity, accuracy, and relevance, selecting the most insightful representation. This method demonstrably enhances the LLM's capacity to generate meaningful, musician-relevant insights from complex musical data

\subsection{Music Production-Oriented Feedback Instruction}

We devise advanced prompting techniques, synthesizing established engineering approaches with novel methodologies, to transform the LLM into a sophisticated, user-centric music assessment system capable of generating high-caliber analysis and feedback.

\subsubsection{Foundational Prompt Template.}

We structure our prompt into three distinct parts: primary function, track scoring process, and track improvement suggestion. In the primary function component, we provide role-specific instructions, ensuring the LLM understands its role as a music feedback provider. We design the track scoring process to avoid excessive praise and identify areas needing improvement. To guide the LLM in assigning scores, we introduce predefined categories such as \emph{Creativity and Originality}, \emph{Genre Fidelity}, \emph{Conveyability}, \emph{Musical Richness}, and \emph{Track Memorability}. This approach addresses LLMs' tendency to offer only positive assessments, which often lack critical suggestions for music improvement. Finally, we include a track improvement suggestion section, guiding the LLM through a process to provide meaningful feedback. We will elaborate on the details of this track improvement suggestion process in the following section

\subsubsection{Track Improvement Suggestion}

We apply several prompt engineering techniques and self-designed strategies to ensure the LLM's music understanding is meaningful to musicians.

\begin{figure}[h]
    \centering
    \includegraphics[width=\linewidth]{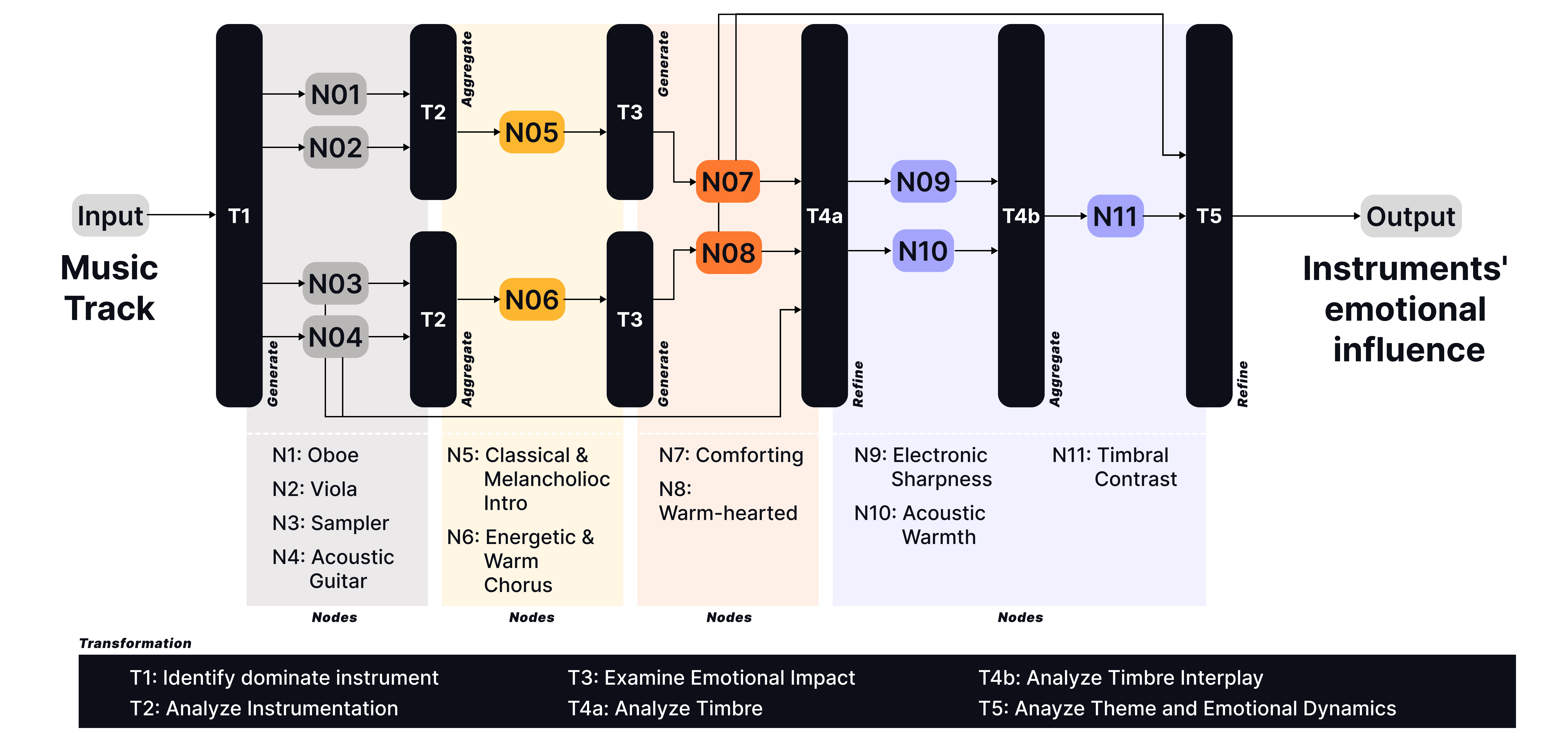}
    \caption{Graph of Thoughts (GoT) approach applies to analyze dominant instruments' impact on a track's emotional tone. Our workflow progresses through generate (T1, T3), aggregate (T2, T4b), and refine (T4a, T5) transformations, adhering to the original GoT framework. We represent analytical steps as nodes (N1-N11), demonstrating how GoT decomposes complex musical analysis into interconnected nodes. This structure elucidates how instrument interactions contribute to the track's overall emotional impact. Through this application of GoT, we enable a systematic exploration of the relationship between instrumental composition and emotional resonance in music.}
    \label{fig:got}
\end{figure}

\textbf{Graph-of-Thought (GoT).} The GoT prompting technique has demonstrated significant potential in enhancing LLMs' reasoning capabilities \cite{besta2024graph}. In our study, we apply this approach to the complex domain of music production by deconstructing high-level thinking processes into several key factors. These factors encompass both objective and subjective elements of musical composition and analysis. We consider technical aspects such as instruments, rhythm, and timbre for the objective side. To complement these, we incorporate subjective elements, including music's emotion and theme. By systematically integrating these objective and subjective dimensions, we enable the LLM to interpret and analyze musical pieces with greater accuracy, thus bridging the gap between technical analysis and artistic interpretation. We present an illustrative example in Figure \ref{fig:got}.

\textbf{Feedback Mechanism.} Our feedback mechanism is designed to enhance the interaction between LLM and musicians, focusing on providing actionable insights and maintaining user engagement. We integrate data-driven analysis with contextual understanding, addressing key aspects of musical composition such as melody, harmony, rhythm, and production techniques. Through iterative testing, we refine the instruction set to mitigate the LLM's tendency towards shallow data parroting, instead promoting critical thinking and deeper analysis. Our system employs clear, constructive language to deliver balanced critiques, ensuring feedback is both informative and motivating. To enhance user experience, we implement a conversational tone reminiscent of a friendly rapper, utilizing metaphors and analogies to elucidate complex musical concepts. This strategy is coupled with the integration of tailored questions, significantly improving user engagement and facilitating a more dynamic, user-centered interaction. Our research demonstrates that this refined feedback mechanism not only provides musicians with practical, applicable advice but also fosters a more engaging and productive dialogue between the AI system and its users.

\textbf{Person Switching. } The person-switching strategy is incorporated in our instruction design to optimize the LLM's performance in music evaluation tasks. We strategically employ different grammatical persons to delineate various aspects of the LLM's operation. We utilize the first-person perspective to foster an internalized evaluation process, encouraging the LLM to adopt a more engaged and personal stance. This technique aims to enhance the depth of analysis and promote active reasoning. Concurrently, we employ second-person constructions when presenting objective information or factual content, creating a clear demarcation between the LLM's internal reasoning processes and external reference points. This distinction aids in maintaining accuracy and proper attribution in the LLM's responses. We use imperative sentences to convey non-negotiable rules and guidelines, ensuring strict adherence to the established evaluation framework and maintaining the integrity of the assessment process. Our person-switching methodology effectively differentiates between the LLM's thought processes, factual knowledge, and operational constraints, resulting in more nuanced and contextually appropriate responses in music evaluation tasks. By carefully balancing these linguistic approaches, we create a comprehensive instruction set that guides the LLM to produce evaluations that are both insightful and aligned with predefined standards.

\subsection{Web-based UI}

The web application is built using the Gradio \cite{abid2019gradio} framework. The typical workflow involves users uploading audio files or pasting YouTube links for analysis. Our system conducts various analyses, generates a music report, and provides it to the LLM. The LLM then offers initial scoring and improvement suggestions in the chat window. Users can pose follow-up questions through the text input interface, facilitating an interactive, user-driven analysis process that transforms how independent producers receive personalized feedback on their work.

\begin{figure}[h]
\centering
\includegraphics[width=\linewidth]{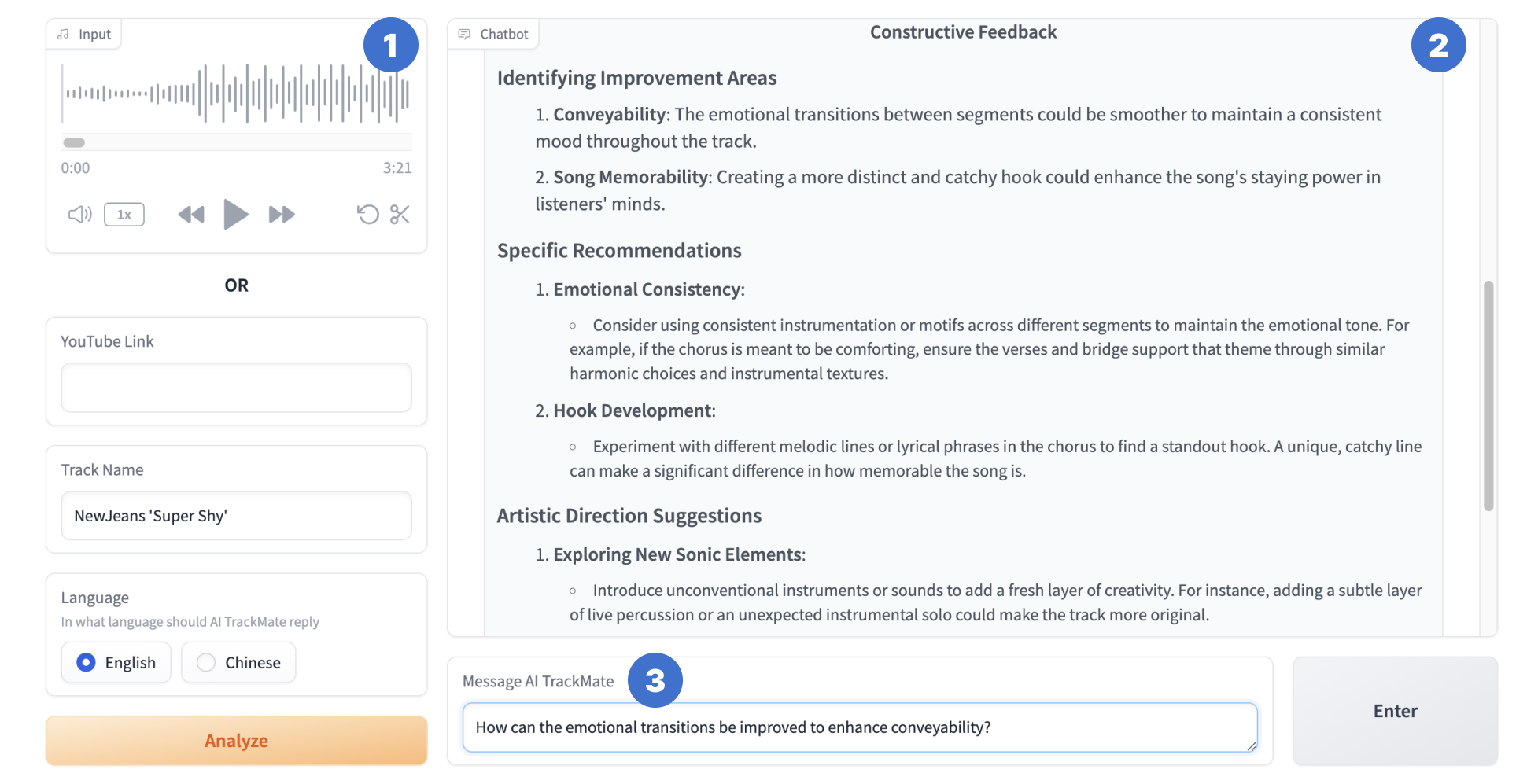}
\caption{The user interface consists of: (1) \textbf{Audio input components} for file upload or YouTube link input. (2) \textbf{A chat window} displaying the LLM's scoring and suggestions. (3) \textbf{A text input} for user questions and LLM responses.}
\label{fig:webui}
\end{figure}

\section{Pilot Study}
\label{pilot_study}

We conducted an exploratory pilot study with a music producer to gather initial insights about our system's approach and potential impact. The study consisted of a questionnaire, an onboarding session, an AI-assisted track analysis, and a recorded conversation, concluding with a semi-structured interview. Our analysis synthesizes observations from the producer profile, AI-producer dialogue, and interview feedback to understand how the system performs in a real-world scenario.

\subsection{Producer Profiles}

Our pilot study participant was a 24-year-old self-taught producer with four years of experience, representing an example of the "bedroom producer" demographic. His experience provided insight into the challenges of skill development and self-evaluation in isolated environments. When discussing peer feedback, he noted its subjective limitations, stating: \textit{``Some people's suggestions are completely based on their own preferences, which only have a little reference value for me''}. This individual case suggests potential value in more objective, intention-aligned feedback systems. The participant's high comfort with music technology (10/10 rating for AI tools) indicated readiness to engage with AI-assisted systems, though broader studies would be needed to assess general producer attitudes.

\subsection{AI-Producer Dialogues}

The AI-Producer dialogue provided several interesting observations about interaction patterns and system capabilities. The conversation naturally progressed from general queries (\textit{``Why are my lyrics and melody disconnected?''}) to specific technical discussions (\textit{``How should I modify my chord progression?''}),  indicating a progression toward more focused technical inquiries. This evolution suggests potential value in facilitating structured technical discussions for solitary producers.

The dialogue culminated with the producer asking, \textit{``If modifications were made based on your suggestions, how would the emotional flow of the track be after these changes? Can you help me compare the emotional analysis of the modified track with the unmodified original version?''} In response, the AI provided a comparative analysis: \textit{``The existing version has relatively smooth emotional changes between verse and chorus, but lacks clear emotional highs and lows. The modified version maintains overall warmth consistency while introducing richer chord changes, giving the chorus section more emotional layers and dynamism''}. This example shows how the system attempts to link technical modifications with their potential emotional impact, illustrating an approach to integrating technical and perceptual feedback in music production.

Analysis of the dialogue content showed that 75\% of AI responses combined technical suggestions with emotional/perceptual feedback. This observation suggests a possible framework for balancing technical and artistic elements in AI-assisted music production feedback.

\subsection{Producer Opinion on AI Feedback}

Initial feedback from our pilot study participant offered exploratory insights into the system's current capabilities and limitations. In examining the tool's analysis approach, the participant noted:  \textit{``It considers aesthetic aspects from a technical perspective and thinks about different parts in a comprehensive way''}, suggesting one possible method for connecting technical and aesthetic feedback. When discussing comparative experiences with peer feedback, they observed that the system \textit{``gradually delves into the details that we music creators care about''}, indicating potential advantages of structured analytical approaches. The participant also noted the system's handling of emotional themes and instrumentation analysis.

Through the exploratory session, several observations emerged about the system's current implementation. The participant found the feedback structure helpful for their creative process, noting potential applications for skill development. For novice musicians in particular, they suggested the system might offer guidance on conceptual aspects not commonly found in general tutorials. The feedback session also revealed several areas for potential enhancement, including more detailed mixing analysis, better DAW integration, improved lyrics and vocal analysis, and comparison capabilities with reference tracks. A key suggestion focused on workflow integration, specifically the possibility of in-DAW feedback during the production process.

While these observations from a single user study cannot be generalized, they provide useful directions for future investigation. The feedback highlights both promising aspects and areas needing refinement, offering specific technical and conceptual considerations for future development of AI-assisted music production tools.

\section{Conclusion \& Discussion}

In this paper, we propose AI TrackMate, an innovative LLM-based music chatbot for production-oriented feedback. Our system leverages LLMs' inherent musical knowledge, integrating a Music Analysis Module, LLM Readable Music Report, and Music Production-Oriented Feedback Instructions. The latter employs advanced techniques like Graph-of-Thought prompting to guide the LLM in thinking and responding like an experienced music producer, providing comprehensive, tailored feedback for bedroom producers.

Our current implementation has several inherent limitations that point to future work directions. The system primarily focuses on conventional music structures, suggesting the need for expanded genre coverage including classical and experimental music. Technical constraints in mixing analysis, vocal interpretation, and lyrical content analysis could be addressed through enhanced capabilities. Additionally, moving from the current per-submission model to real-time analysis could enable better integration with active production workflows. Our exploratory pilot study provided initial insights into these aspects, and we leave comprehensive evaluation studies across various genres and skill levels as important directions for future research.

\bibliographystyle{plainnat}
\bibliography{reference}

\end{document}